\documentclass[conference,a4paper]{IEEEtran}
\IEEEoverridecommandlockouts
\usepackage{cite}
\usepackage{amsmath,amssymb,amsfonts}
\usepackage{graphicx}
\usepackage{textcomp}
\usepackage{xcolor}
\usepackage{svg}
\usepackage{siunitx}
\usepackage{xinttools}
\usepackage{tikz}
\usetikzlibrary{positioning}
\usepackage{mathtools}
\usepackage{neuralnetwork}
\usepackage{multicol}
\usepackage{multirow}
\usepackage{lipsum}

\usepackage{amsmath}
\usepackage{amsfonts}
\usepackage{dsfont}
\usepackage{tikz}
\usepackage{mathdots}
\usepackage{yhmath}
\usepackage{cancel}
\usepackage{color}
\usepackage{siunitx}
\usepackage{array}
\usepackage{multirow}
\usepackage{amssymb}
\usepackage{gensymb}
\usepackage{tabularx}
\usepackage{extarrows}
\usepackage{booktabs}
\usetikzlibrary{fadings}
\usetikzlibrary{patterns}
\usetikzlibrary{shadows.blur}
\usetikzlibrary{shapes}
\usepackage[symbol]{footmisc}
\newcommand{\code}[1]{\texttt{#1}}

\makeatletter

\def\BibTeX{{\rm B\kern-.05em{\sc i\kern-.025em b}\kern-.08em
    T\kern-.1667em\lower.7ex\hbox{E}\kern-.125emX}}
\begin{document}

\makeatletter 
\newcommand{\linebreakand}{%
  \end{@IEEEauthorhalign}
  \hfill\mbox{}\par
  \mbox{}\hfill\begin{@IEEEauthorhalign}
}
\makeatother

\title{Parameter Estimation in Electrical Distribution Systems with limited Measurements using Regression Methods}

\author{\IEEEauthorblockN{Steven de Jongh}
\IEEEauthorblockA{\small \textit{Karlsruhe Institute of Technology (KIT)}\\ \normalsize
Karlsruhe, Germany \\
steven.dejongh@kit.edu}
\and
\IEEEauthorblockN{Felicitas Mueller}
\IEEEauthorblockA{\small \textit{Karlsruhe Institute of Technology (KIT) }\\ \normalsize
Karlsruhe, Germany \\
felicitas.mueller@kit.edu}
\and
\IEEEauthorblockN{Claudio A. Ca\~nizares}
\IEEEauthorblockA{\small \textit{University of Waterloo} \normalsize\\
Waterloo, Ontario, Canada \\
ccanizares@uwaterloo.ca}
\and
\linebreakand
\IEEEauthorblockN{Thomas Leibfried}
\IEEEauthorblockA{\small \textit{Karlsruhe Institute of Technology (KIT)} \\ \normalsize
Karlsruhe, Germany \\
thomas.leibfried@kit.edu}
\and 
\IEEEauthorblockN{Kankar Bhattacharya}
\IEEEauthorblockA{\small \textit{University of Waterloo} \normalsize \\
Waterloo, Ontario, Canada \\
kankar@uwaterloo.ca}}

\renewcommand{\thefootnote}{\fnsymbol{footnote}}
\newcommand\blfootnote[1]{%
  \begingroup
  \renewcommand\thefootnote{}\footnote{#1}%
  \addtocounter{footnote}{-1}%
  \endgroup}

\maketitle

\begin{abstract}
This paper presents novel methods for parameter identification in electrical grids with small numbers of spatially distributed measuring devices, which is an issue for distribution system operators managing aged and not properly mapped underground Low Voltage (LV) grids, especially in Germany. For this purpose, the total impedance of individual branches of the overall system is estimated by measuring currents and voltages at a subset of all system nodes over time. It is shown that, under common assumptions for electrical distsribution systems, an estimate of the total impedance can be made using readily computable proxies. Different regression methods are then used and compared to estimate the total impedance of the respective branches, with varying weights of the input data. The results on realistic LV feeders with different branch lengths and number of unmeasured segments are discussed and multiple influencing factors are investigated through simulations. It is shown that estimates of the total impedances can be obtained with acceptable quality under realistic assumptions.
\end{abstract}


\section{Introduction}

\blfootnote{*This work was funded by Mitacs, NSERC and KIT. \\ The A.I. based language models DeepL, OpenAI GPT-3.5 and you.com were used for text editing and other language-related tasks.}
Access to precise mathematical models is paramount in facilitating the practical application of grid calculation algorithms that enable the computation of grid congestion, prediction of future scenarios, and exploration of grid expansion options. These models can be accurately derived from specific variables, such as line parameters, known line models, and line lengths. However, in reality, such data may not be readily available or may contain errors as in the case of aged underground LV grids that are not properly mapped, which is a problem currently being faced by German distribution system operators, resulting in a challenging grid identification problem. To address this issue, Parameter Identification (PI) methods can be employed to deduce model parameters from system measurements. As the number of measurement sites in the grid is typically limited, especially in LV distribution system, parameter identification methods become increasingly important for grids with low sensor penetration.

The technical literature contains numerous publications that address PI in electrical grids, which is a subset of system identification, that is utilized in electrical grids for identifying component parameters \cite{param1, param2}, and topology identification \cite{topologyoverview, topology2}. In many applications, these methods are integrated into a unified framework that seeks to identify both correct topology and parameters as a single task, as seen in \cite{bayesian}. However, the proposed methodologies often assume the availability of numerous and expensive sensors, such as phasor measurement units, for instance, \cite{OnIdentification} and \cite{dejongh} assume measurements of all complex voltages and nodal currents. In reality, the number of available measurement units is limited, particularly in electrical distribution grids, especially for LV networks, rendering measurements at all nodes unfeasible. For example, \cite{kronest} considers such a scenario and employs Kron-reduction to perform identification based on a reduced graph. In \cite{marulli}, a grid with limited measurement penetration is considered and a penetration of more than 80\% is proposed for acceptable results. Although the presented methods can deal with sparse measurement setups, they do not provide realistic results for PI in distribution grids, where often only a much smaller fraction (much less than 80\%) of spatially distributed measurement points is available. Therefore, in contrast to the existing literature, this paper addresses PI in electrical grids with sparse available sensors. By utilizing regression methods, the total impedance of the respective branches is estimated and the influence of number of unmeasured segments and measurement data resolution is investigated.

The paper is structured as follows: In Section \ref{sec:methodology}, the modelling of unmeasured branches in electrical grids is introduced, applying asymptotic analysis for the formulation of bounds on the total impedances, as well as regression methods that allow the estimation of total impedance of the respective branches. Section \ref{sec:results} discusses the results of the proposed methods based on realistic LV feeders, and relevant conclusions as well as a future outlook are given in Section \ref{sec:outlook}.

\section{Methodology}
\label{sec:methodology}
\subsection{Modelling of Electrical Grids}
Electrical grids can be represented as mathematical graphs $\mathcal{G}$, composed of a set of nodes $\mathcal{N}=\{1,2,\dots,n_{\mathrm{nodes}}\}$ and a set of edges $\mathcal{E}=\{ (i,j) \; | \; i, j \in \mathcal{N} \}$. Each edge in the graph has an associated impedance value, which can be used to parameterize the grid model. When all system parameters are known, it is possible to derive sparse matrices, such as the node admittance matrix. These matrices can then be utilized to calculate the power flows in the grid, based on the known nodal consumption. The edges in graph $\mathcal{G}$ are characterized by their impedances, which define the overall system properties and determine the power and current flows in the system.

\subsection{Branch Modelling}
In this study, the concept of a \textit{measured branch} is applied, which is a connection between two nodes, denoted as '$\mathrm{in}$' and '$\mathrm{out}$', comprising $K$ individual impedances or \textit{segments}. Each segment is interspersed with an unmeasured load, which results in $K-1$ unmeasured loads. As illustrated in Fig. \ref{fig:branchmodel}, it is assumed that voltage measurements are available at both the input $v_{\mathrm{in}}$ and output $v_{\mathrm{out}}$ of the branch. Furthermore, both the inflowing current $i_{\mathrm{in}}$ and outflowing current $i_{\mathrm{out}}$ are measured. The measurements are taken at discrete time intervals over a prolonged period of time $T$ and can be represented as vectors, denoted as $\mathbf{v}_{\mathrm{in}} = [v_{\mathrm{in}}(t_1), v_{\mathrm{in}}(t_2), \dots, v_{\mathrm{in}}(T)]$ and $\mathbf{i}_{\mathrm{in}} = [i_{\mathrm{in}}(t_1), i_{\mathrm{in}}(t_2), \dots, i_{\mathrm{in}}(T)]$ for the measured quantities at the output of the branch.

\tikzset{every picture/.style={line width=0.75pt}}
\begin{figure}[ht]
    \begin{tikzpicture}[x=0.75pt,y=0.75pt,yscale=-1.2,xscale=1.2]
    
    \draw  [draw opacity=0][fill={rgb, 255:red, 155; green, 155; blue, 155 }  ,fill opacity=0.1 ] (70.67,48.29) .. controls (70.67,40.33) and (77.13,33.87) .. (85.09,33.87) -- (259.82,33.87) .. controls (267.79,33.87) and (274.24,40.33) .. (274.24,48.29) -- (274.24,91.57) .. controls (274.24,99.54) and (267.79,106) .. (259.82,106) -- (85.09,106) .. controls (77.13,106) and (70.67,99.54) .. (70.67,91.57) -- cycle ;
    \draw   (55.57,63.39) .. controls (55.57,61.12) and (57.41,59.29) .. (59.68,59.29) .. controls (61.95,59.29) and (63.78,61.12) .. (63.78,63.39) .. controls (63.78,65.66) and (61.95,67.5) .. (59.68,67.5) .. controls (57.41,67.5) and (55.57,65.66) .. (55.57,63.39) -- cycle ;
    \draw   (75.33,54.67) -- (104.78,54.67) -- (104.78,72.02) -- (75.33,72.02) -- cycle ;
    \draw    (63.78,63.39) -- (75.39,63.27) ;
    \draw    (104.24,63.27) -- (120.57,63.06) ;
    \draw   (141.33,54.67) -- (170.78,54.67) -- (170.78,72.02) -- (141.33,72.02) -- cycle ;
    \draw    (128.78,63.06) -- (141.19,63.1) ;
    \draw    (174.49,63.53) -- (186.33,63.39) ;
    \draw   (242,55) -- (271.45,55) -- (271.45,72.35) -- (242,72.35) -- cycle ;
    \draw    (271.16,63.87) -- (283,63.73) ;
    \draw    (231.12,63.73) -- (242.73,63.6) ;
    \draw  [fill={rgb, 255:red, 0; green, 0; blue, 0 }  ,fill opacity=1 ] (200.53,62.79) .. controls (200.53,61.99) and (201.19,61.33) .. (201.99,61.33) .. controls (202.8,61.33) and (203.45,61.99) .. (203.45,62.79) .. controls (203.45,63.6) and (202.8,64.25) .. (201.99,64.25) .. controls (201.19,64.25) and (200.53,63.6) .. (200.53,62.79) -- cycle ;
    \draw  [fill={rgb, 255:red, 0; green, 0; blue, 0 }  ,fill opacity=1 ] (207.2,62.88) .. controls (207.2,62.07) and (207.85,61.42) .. (208.66,61.42) .. controls (209.46,61.42) and (210.12,62.07) .. (210.12,62.88) .. controls (210.12,63.68) and (209.46,64.33) .. (208.66,64.33) .. controls (207.85,64.33) and (207.2,63.68) .. (207.2,62.88) -- cycle ;
    \draw  [fill={rgb, 255:red, 0; green, 0; blue, 0 }  ,fill opacity=1 ] (213.95,62.88) .. controls (213.95,62.07) and (214.6,61.42) .. (215.41,61.42) .. controls (216.21,61.42) and (216.87,62.07) .. (216.87,62.88) .. controls (216.87,63.68) and (216.21,64.33) .. (215.41,64.33) .. controls (214.6,64.33) and (213.95,63.68) .. (213.95,62.88) -- cycle ;
    \draw   (120.57,63.06) .. controls (120.57,60.79) and (122.41,58.95) .. (124.68,58.95) .. controls (126.95,58.95) and (128.78,60.79) .. (128.78,63.06) .. controls (128.78,65.33) and (126.95,67.16) .. (124.68,67.16) .. controls (122.41,67.16) and (120.57,65.33) .. (120.57,63.06) -- cycle ;
    \draw   (186.33,63.39) .. controls (186.33,61.12) and (188.17,59.29) .. (190.44,59.29) .. controls (192.71,59.29) and (194.54,61.12) .. (194.54,63.39) .. controls (194.54,65.66) and (192.71,67.5) .. (190.44,67.5) .. controls (188.17,67.5) and (186.33,65.66) .. (186.33,63.39) -- cycle ;
    \draw   (222.91,63.73) .. controls (222.91,61.46) and (224.74,59.62) .. (227.01,59.62) .. controls (229.28,59.62) and (231.12,61.46) .. (231.12,63.73) .. controls (231.12,65.99) and (229.28,67.83) .. (227.01,67.83) .. controls (224.74,67.83) and (222.91,65.99) .. (222.91,63.73) -- cycle ;
    \draw   (283,63.73) .. controls (283,61.46) and (284.84,59.62) .. (287.11,59.62) .. controls (289.37,59.62) and (291.21,61.46) .. (291.21,63.73) .. controls (291.21,65.99) and (289.37,67.83) .. (287.11,67.83) .. controls (284.84,67.83) and (283,65.99) .. (283,63.73) -- cycle ;
    \draw    (124.68,67.16) -- (124.59,83.8) ;
    \draw [shift={(124.58,86.8)}, rotate = 270.29] [fill={rgb, 255:red, 0; green, 0; blue, 0 }  ][line width=0.08]  [draw opacity=0] (8.93,-4.29) -- (0,0) -- (8.93,4.29) -- cycle    ;
    \draw    (190.44,67.5) -- (190.35,84.13) ;
    \draw [shift={(190.34,87.13)}, rotate = 270.29] [fill={rgb, 255:red, 0; green, 0; blue, 0 }  ][line width=0.08]  [draw opacity=0] (8.93,-4.29) -- (0,0) -- (8.93,4.29) -- cycle    ;
    \draw    (227.01,67.83) -- (226.93,84.47) ;
    \draw [shift={(226.91,87.47)}, rotate = 270.29] [fill={rgb, 255:red, 0; green, 0; blue, 0 }  ][line width=0.08]  [draw opacity=0] (8.93,-4.29) -- (0,0) -- (8.93,4.29) -- cycle    ;
    \draw    (46.91,73.2) -- (63.24,73.48) ;
    \draw [shift={(66.24,73.53)}, rotate = 180.99] [fill={rgb, 255:red, 0; green, 0; blue, 0 }  ][line width=0.08]  [draw opacity=0] (10.72,-5.15) -- (0,0) -- (10.72,5.15) -- (7.12,0) -- cycle    ;
    \draw    (279.91,73.2) -- (296.24,73.48) ;
    \draw [shift={(299.24,73.53)}, rotate = 180.99] [fill={rgb, 255:red, 0; green, 0; blue, 0 }  ][line width=0.08]  [draw opacity=0] (10.72,-5.15) -- (0,0) -- (10.72,5.15) -- (7.12,0) -- cycle    ;
    \draw    (105.34,62.73) -- (117.57,62.99) ;
    \draw [shift={(120.57,63.06)}, rotate = 181.25] [fill={rgb, 255:red, 0; green, 0; blue, 0 }  ][line width=0.08]  [draw opacity=0] (10.72,-5.15) -- (0,0) -- (10.72,5.15) -- (7.12,0) -- cycle    ;
    \draw    (170.71,63.34) -- (183.33,63.38) ;
    \draw [shift={(186.33,63.39)}, rotate = 180.18] [fill={rgb, 255:red, 0; green, 0; blue, 0 }  ][line width=0.08]  [draw opacity=0] (10.72,-5.15) -- (0,0) -- (10.72,5.15) -- (7.12,0) -- cycle    ;
    \draw    (60.5,34) .. controls (111.2,15.53) and (240.59,22.2) .. (286.35,35.92) ;
    \draw [shift={(287.71,36.34)}, rotate = 197.46] [color={rgb, 255:red, 0; green, 0; blue, 0 }  ][line width=0.75]    (10.93,-3.29) .. controls (6.95,-1.4) and (3.31,-0.3) .. (0,0) .. controls (3.31,0.3) and (6.95,1.4) .. (10.93,3.29)   ;
    
    \draw (82,42) node [anchor=north west][inner sep=0.75pt]    {$z_{1}$};
    \draw (148.33,42) node [anchor=north west][inner sep=0.75pt]    {$z_{2}$};
    \draw (247.67,42) node [anchor=north west][inner sep=0.75pt]    {$z_{K}$};
    \draw (44.67,40.54) node [anchor=north west][inner sep=0.75pt]    {$v_{\mathrm{in}}$};
    \draw (277.33,40.54) node [anchor=north west][inner sep=0.75pt]    {$v_{\mathrm{out}}$};
    \draw (45.33,74.88) node [anchor=north west][inner sep=0.75pt]    {$i_{\mathrm{in}}$};
    \draw (279,74.54) node [anchor=north west][inner sep=0.75pt]    {$i_{\mathrm{out}}$};
    \draw (118,86.54) node [anchor=north west][inner sep=0.75pt]  [font=\scriptsize]  {$\tilde{i}_{1}$};
    \draw (183,85.54) node [anchor=north west][inner sep=0.75pt]  [font=\scriptsize]  {$\tilde{i}_{2}$};
    \draw (220.33,85.21) node [anchor=north west][inner sep=0.75pt]  [font=\scriptsize]  {$\tilde{i}_{K-1}$};
    \draw (110,45.9) node [anchor=north west][inner sep=0.75pt]  [font=\scriptsize]  {$i_{1}$};
    \draw (175,45.9) node [anchor=north west][inner sep=0.75pt]  [font=\scriptsize]  {$i_{2}$};
    \draw (155.5,6.4) node [anchor=north west][inner sep=0.75pt]    {$\Delta v$};
    \end{tikzpicture}
    \caption{Branch model with unmeasured nodes between in- and out-nodes}
    \label{fig:branchmodel}
\end{figure}
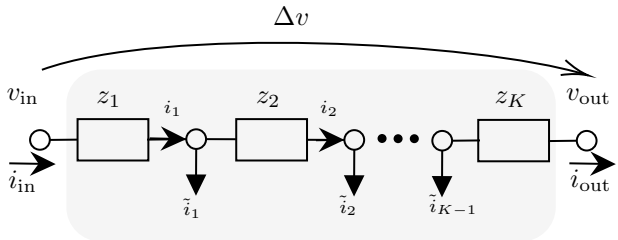

The voltage drop across each branch depicted in Fig. \ref{fig:branchmodel} at discrete time steps can be mathematically formulated as follows, wherein the tilde-annotated variables denote unmeasured quantities and $i_k$ denotes the current flowing through the impedance within segment $k$:

\begin{equation}
    \Delta v = v_{\mathrm{in}} - v_{\mathrm{out}} = \sum_{k=1}^{K} z_k \cdot i_k \\ 
    \label{eq:voltagedrop}
\end{equation}

\begin{equation}
    i_k = i_{\mathrm{in}}-\sum_{m=1}^{k-1}\tilde{i}_m
    \label{eq:intermediatecurrents}
\end{equation}

If only one unmeasured node exists between the input and output nodes (i.e., $K=2$), the system of equations remains determinable, enabling the derivation of the unmeasured current $\tilde{i}_1$ by applying Kirchhoff's current law. Here, the sought-after parameters $z_1$ and $z_2$ can be deduced by solving \eqref{eq:voltagedrop}, or alternatively, by utilizing the least squares approach in case of the presence of measurements of multiple time steps. However, when the number of segments exceeds $K\geq2$, the impedances cannot be uniquely solved for, needing the utilization of multiple time-step measurements to arrive at a conclusive estimation of the total impedance of the branch.

\subsection{Asymptotic analysis}
\label{subsec:limitconsiderations}
The following assumptions are made, which enable the estimation of the total impedance via regression techniques.

\begin{itemize}
    \item \textit{Assumption 1:} The loads, denoted by $\tilde{i}_k$, for all $k \in \{1, 2, \dots, K-1\}$, are strictly positive, and there is no current injection back into the branch.
    \item \textit{Assumption 2:} The voltage decreases along the branch, and the current flowing into the branch, denoted by $i_{\mathrm{in}}$, is greater than or equal to the current flowing out of the branch $i_{\mathrm{out}}$, leading to $i_{\mathrm{in}} \geq i_{\mathrm{out}}$.
\end{itemize}
These assumptions align with typical assumptions made in passive electrical distribution grids. In the event that Assumption 1 is not met, the data pertaining to the respective time step may be disregarded. Conversely, if Assumption 1 is satisfied, and the branch experiences single-sided supply (e.g., in a radial grid), Assumption 2 is automatically satisfied. Note that in the case of photovoltaic systems, night measurements may be used to satisfy Assumption 2.

The following equation represents the \textit{participation factor}, denoted by $f$, which characterizes the ratio between inflowing and outflowing currents from the branch, with $0 \leq f \leq 1$ under Assumptions 1 and 2.

\begin{equation}
    f = \frac{i_{\mathrm{out}}}{i_{\mathrm{in}}} \qquad \text{with} \; i_{\mathrm{in}} > 0
    \label{eq:participationfactor}
\end{equation}

Two asymptotic bounds can be derived from \eqref{eq:participationfactor}. Thus, in the case when all intermediate load currents are zero and $i_{\mathrm{in}} \neq 0$, the participation factor approaches one, i.e., $f \rightarrow 1$. In this case, the total impedance can be determined by substituting all $i_k$ in \eqref{eq:voltagedrop} with $i_{\mathrm{in}}$ and solving for $\sum_{k=1}^{K} z_k$. Furthermore, an upper bound $z_{\mathrm{tot}, \mathrm{ub}}$ and lower bound $z_{\mathrm{tot}, \mathrm{lb}}$ can be  formulated for the total impedance as follows.

\begin{equation}
    z_{\mathrm{tot}, \mathrm{lb}} = \frac{\Delta v}{i_{\mathrm{in}}};\qquad z_{\mathrm{tot}, \mathrm{ub}} = \frac{\Delta v}{i_{\mathrm{out}}}
    \label{eq:lowerupper}
\end{equation}
Note that these bounds can be calculated at each time step, and the bound with the overall tightest constraint gives the valid parameter range for $z_{\mathrm{tot}}$.

\subsection{Total Impedance Estimation}
Several techniques to estimate the total impedance of branches with varying numbers of segments, denoted by $K$, are proposed here. Although obtaining the actual total impedance requires $f\rightarrow 1$, such a condition may not occur in practice, since all load currents are rarely zero simultaneously. To address this issue, regression techniques can be used to approximate the estimated total impedance, denoted by $\hat{z}_{\mathrm{tot}}$, as a function of the participation factor $f$. Specifically, the lower bound $z_{\mathrm{tot},\mathrm{lb}}$ can be estimated from the relationship between $f$ and the estimated total impedance by evaluating the regression line at the point where $f=1.0$. For this purpose, the measurements at all time steps are used to create the stacked column vectors $\mathbf{f} = [f(t_1), f(t_2), \dots f(T)]^\intercal$, $\Delta \mathbf{v} = [\Delta v(t_1), \Delta v(t_2), \dots, \Delta v(T)]^\intercal$ and $\mathbf{i}_{\mathrm{in}} = [i_{\mathrm{in}}(t_1), i_{\mathrm{in}}(t_2), \dots, i_{\mathrm{in}}(T)]^\intercal$ to calculate the vector of lower bounds $z_{\mathrm{tot}, \mathrm{lb},i}=\frac{\Delta v_i}{i_{\mathrm{in},i}} \quad \forall i$. Then the following least squares problem can be solved to obtain the slope $\beta_0$ and intercept $\beta_1$:

\begin{equation}
    \underset{\beta_0, \beta_1}{\mathrm{min}} \qquad \left\| \mathbf{z}_{\mathrm{tot},\mathrm{lb}} - \big[\mathbf{f} \quad \mathds{1}_{T\times 1}\big] \cdot \begin{bmatrix} \beta_0 \\ \beta_1 \end{bmatrix} \right\|^2    
    \label{eq:leastsquares}
\end{equation}
Finally, an estimate of the total impedance can be obtained as follows.

\begin{equation}
     \hat{z}_{\mathrm{tot}} = \beta_0 \cdot 1 + \beta_1
    \label{eq:estimate}
\end{equation}

The following approaches to find estimates of the impedances are compared next:
\begin{itemize}
    \item $\mathrm{lin}$: The estimate is obtained using $\max(\hat{z}_{\mathrm{tot}}, \max(\mathbf{z}_{\mathrm{tot}, \mathrm{lb}}))$.
    \item $\mathrm{lin}_w$: The estimate is obtained by solving a Weighted Least Squares (WLS) version of \eqref{eq:leastsquares} with weights $w(t) = \mathbf{f}(t)$ to weigh samples closer to $f=1$ more heavily.
    \item $\mathrm{lin}_{w^2}$: WLS is applied with a quadratic weight $w(t) = \mathbf{f}(t)^2$.
    \item $\mathrm{mean}_{\mathrm{lb},\mathrm{ub}}$: The estimate is the mean between the lowest upper bound and highest lower bound $\hat{z}_{\mathrm{tot}} = \max(\mathbf{z}_{\mathrm{tot},\mathrm{lb}}) + \frac{\min(\mathbf{z}_{\mathrm{tot},\mathrm{ub}}) - \max(\mathbf{z}_{\mathrm{tot},\mathrm{lb}})}{2}$.
\end{itemize}
The first three methods are based on a linear regression that can be performed either without weighting factors $\mathrm{lin}$, or with weighting factors $\mathrm{lin}_w$ and $\mathrm{lin}_{w^2}$. In this case, the weighting takes place based on the participation factors, such that data samples with higher participation factors are more heavily weighted, so that the estimated impedances get closer to the actual impedance. The $\mathrm{mean}_{\mathrm{lb}, \mathrm{ub}}$ method represents a straightforward method of estimation using the upper and lower bounds. In situations with high participation factors, this can serve as a first approximation of the total impedance.

\section{Results}
\label{sec:results}
\subsection{Simulation Setup}
\label{sec:simulationsetup}
To validate the methods introduced in Section \ref{sec:methodology}, branches with different numbers $K$ of segments are considered as single phase LV grids. Power-flow simulations are performed using the \code{pandapower} package in \code{python}  for each individual time step in the data \cite{pandapower}. Profiles are randomly drawn from the household profiles provided in the IEEE European LV Test Feeder, in \cite{ieeeprofiles}, with $\Delta t = 1 \; \mathrm{min}$ and $T=1440$. The length of all cable segments is defined uniformly random between $l=100 \; \mathrm{m}$ and $l=300 \; \mathrm{m}$. The used cable types are $\mathrm{NAYY 4\times150 mm^2}$, which is a common cable found in German LV grids. In the cases where multiple simulation runs are performed, $n_s$ denotes the number of Monte-Carlo samples that are simulated, where each sample corresponds to a simulation with randomly drawn segment lengths and load profiles.
\subsection{DC circuit Validation}
A DC circuit simulation is used for a branch with $K=3$ to validate the asymptotic bounds derived in Section \ref{subsec:limitconsiderations}. Fig. \ref{fig:DCexample} shows estimation for the lower bound (lb) samples (blue) and for the upper bound (ub) samples (green) for multiple measurements over the respective participation factor $f$, covering the whole range from $f=0.0$ to $f=1.0$. Observe that for $f=1.0$, the samples of the upper and lower bound converge to the true total resistance of the branch. The bounds obtained from the samples, depicted as red lines, limit the possible range of the true total resistance. The relationship for the lower bound is approximately linear and can be adopted during the regression task to identify the total resistance.
\begin{figure}[ht]
    \centering
    \includegraphics[width=\columnwidth]{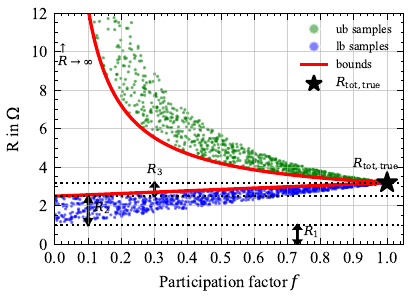}
    \caption{DC circuit with $K=3$ and upper and lower bounds.}
    \label{fig:DCexample}
\end{figure}

\subsection{Scalability and Estimation Quality for AC branches}
\label{sec:scalabilityAC}
In the previous Section, it was demonstrated that the relationship between the participation factor and $z_{\mathrm{tot}, \mathrm{lb}}$ can be utilized to infer the total resistance of a DC circuit through linear regression. In this section, the influence of cable segments $K$ and the regression method used on the estimation error is investigated. To accomplish this, $K$ is varied, as discussed in Section \ref{sec:simulationsetup}, with $n_s=150$ Monte-Carlo simulations being conducted for each method. The quantiles of the estimation error are used to analyze the various identification approaches, where the error is the absolute deviation from the actual total branch impedance, expressed as $\epsilon = \frac{|z_{\mathrm{tot}} - \hat{z}_{\mathrm{tot}}|}{\hat{z}_{\mathrm{tot}}} 100 \%$. 

Fig. \ref{fig:segmentscaling} illustrates the estimation error of methods $\mathrm{lin}$ and $\mathrm{lin}_w$ for an increasing number of segments. Observe that the median error increases for both methods, going from under three percent estimation error at $K=4$ to as high as six percent estimation error for $K=13$, as the increasing number of unmeasured currents $\tilde{i}$ makes it more difficult to estimate the true total impedance. Furthermore, note that the 25\% and 75\% quantiles increase with increasing $K$. The variation in the estimation between different runs is due to the random distribution of segment lengths over multiple runs and the random assignment of load profiles to intermediate nodes in each case. The interquantile range increases for both regression methods as $K$ increases, with the $\mathrm{lin}_w$ method achieving lower estimation errors well below 5\% for $K$ up to 14, showing that more weighting of data points with high participation factor $f$ can be beneficial. The reason for this increase is due to the increase in $f$ range as $K$ increases, as depicted in Fig. \ref{fig:segmentscaling2}, which is to be expected.

\begin{figure}[ht]
    \centering
    \includegraphics[width=\columnwidth]{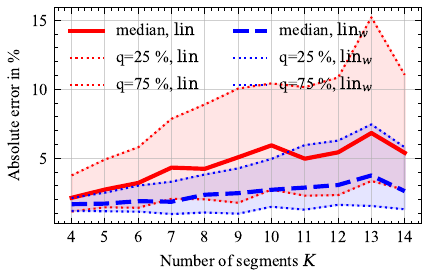}
    \caption{Absolute error of $\hat{z}_{\mathrm{tot}}$ estimation for varying segments $K$ for methods $\mathrm{lin}$ (red) and $\mathrm{lin}_{w}$ (blue).}
    \label{fig:segmentscaling}
\end{figure}

This can be explained by the fact that for smaller numbers of consumers, the proportion of unmeasured consumers that are supplied by the branch is higher in relation to the $K-1$ consumers, and thus the ratio of outgoing flow to incoming flow is accordingly closer to one.

\begin{figure}[ht]
    \centering
    \includegraphics[width=\columnwidth]{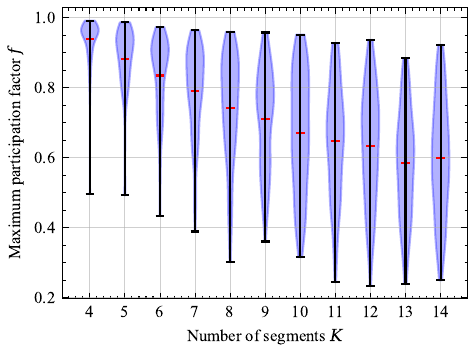}
    \caption{Maximum occuring participation factor for different Monte Carlo samples with respect to number of segments $K$.}
    \label{fig:segmentscaling2}
\end{figure}

\begin{table}[ht]
    \centering
    \caption{Total impedance estimation errors for different PI methods.}
    \begin{tabular}{|c|c||c|c|c|}
         \hline
         & Method & median $\epsilon$ & $75\%$ quant. $\epsilon$ & max $\epsilon$ \\
         \hline
         \multirow{2}{*}{$K=4$} & $\mathrm{mean}_{\mathrm{lb}, \mathrm{ub}}$ & $1.64$ & $\mathbf{1.89}$ & $\mathbf{26.51}$ \\
         \cline{2-5}
         & $\mathrm{lin}$ & $2.13$ & $3.76$ & $53.18$\\
         \cline{2-5}
         \multirow{2}{*}{short branch} & $\mathrm{lin}_{w}$ & $1.67$ & $2.12$ & $53.17$ \\
         \cline{2-5}
         & $\mathrm{lin}_{w^2}$ & $\mathbf{1.62}$ & $1.92$ & $53.17$\\
         \cline{2-5} 
         \hline
         \multirow{2}{*}{$K=14$} & $\mathrm{mean}_{\mathrm{lb}, \mathrm{ub}}$ & $6.58$ & $13.90$ & $68.23$ \\
         \cline{2-5}
         & $\mathrm{lin}$ & $5.39$ & $11.05$ & $66.71$ \\
         \cline{2-5}
         \multirow{2}{*}{long branch} & $\mathrm{lin}_{w}$ & $\mathbf{2.62}$ & $\mathbf{5.79}$ & $\mathbf{56.17}$ \\
         \cline{2-5}
         & $\mathrm{lin}_{w^2}$ & $3.27$ & $6.63$ & $59.87$\\
         \cline{2-5}
         \hline
    \end{tabular}
    \label{tab:estimationerrorsmethods}
\end{table}
Table \ref{tab:estimationerrorsmethods} shows the estimation errors that result from 150 Monte Carlo simulations for the methods presented in this paper. The results for a branch of relatively short length ($K=4$) and long branch ($K=14$) are shown. The method with the best performance, as per the minimum estimation error $\epsilon$, in a specific category is highlighted in bold. Observe that for $K=4$, the median of the estimation error for all methods ranges between $1.62  \%$ and $2.13 \%$ for the short branch. The $\mathrm{lin}_{w^2}$ method provides the most accurate results, followed closely by the $\mathrm{lin}$ and $\mathrm{mean}_{\mathrm{lb},\mathrm{ub}}$ methods. As for the 75\% quantiles and maximum errors, the $\mathrm{mean}_{\mathrm{lb},\mathrm{ub}}$ method yields the best outcomes. It is worth to note that the regression-based methods, including $\mathrm{lin}$, $\mathrm{lin}_w$, and $\mathrm{lin}_{w^2}$, perform better when the data samples with high participation factors are assigned stronger weights. This phenomenon can be explained by the fact that for $K=4$, a considerable amount of data pertaining to high $f$ already exists, rendering simple techniques such as $\mathrm{mean}_{\mathrm{lb},\mathrm{ub}}$ effective, as the limits can accurately estimate the results in this case. Furthermore, note that even though the 75\% quantile exhibits single digit estimation errors for all methods, the maximum estimation errors may reach up to $53.18 \%$. These outliers can cause significant challenges in real-world estimation. In the case of branches with $K=14$, it is evident that the regression methods surpass the relatively simplistic $\mathrm{mean}_{\mathrm{lb},\mathrm{ub}}$ method. Additionally, the results indicate that the estimation errors for the median, 75\% quantile, and maximum values increase as the length of the branch increases. 

The most accurate results in Table \ref{tab:estimationerrorsmethods} for $K=14$ are obtained using regression with $\mathrm{lin}_w$. These results indicate that regression techniques are particularly beneficial for longer branches with more segments, and that assigning weights for data points with higher participation factors, which contain more informative data, yield better outcomes. Observe also that the results obtained using the $\mathrm{mean}_{\mathrm{lb},\mathrm{ub}}$ method deteriorate for longer branches, as the number of data samples with high participation factor decreases, and thus the accuracy of results obtained by considering only the limits is compromised.

\begin{figure}[ht]
    \centering
    \includegraphics[width=\columnwidth]{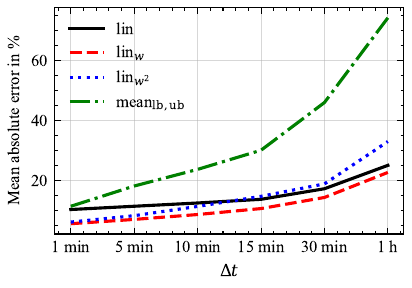}
    \caption{Influence of measurement resolution $\Delta t$ on mean estimation error for $K=14$.}
    \label{fig:influencedt}
\end{figure}

Fig. \ref{fig:influencedt} shows the mean absolute estimation error over $n_s=150$ Monte-Carlo runs for $K=14$ segments and different measurement data resolutions $\Delta t$, for all presented methods. Note that for a resolution of $\Delta t = 1 \; \mathrm{min}$, the lowest mean estimation error is obtained by the method $\mathrm{lin}_{w}$ followed by $\mathrm{lin}_{w^2}$, with errors of $5.61 \%$ and $6.17 \%$, respectively. The error for the $\mathrm{lin}$ and $\mathrm{mean}_{\mathrm{lb}, \mathrm{ub}}$ methods is higher, being $10.35 \%$ and $11.43 \%$, respectively. For all methods, the mean estimation error increases as the measurement data resolution increases, which can be explained by the lower maximum participation factor $f$ that occurs, decreasing from 0.59 at $\Delta t = 1 \; \mathrm{min}$ to 0.26 at $\Delta t = 1 \; \mathrm{h}$. This makes the asymptotic bounds less accurate, leading to lower estimation quality. Two things are apparent here: First, the estimation error of the method $\mathrm{mean}_{\mathrm{lb}, \mathrm{ub}}$ increases the most, since the absence of high participation factor measurements has the greatest effect on the estimate in this case. Second, observe that the methods with weighting (i.e. $w$, $w^2$) show a higher increase than the unweighted regression $\mathrm{lin}$. This suggests that for high measurement data resolutions, unweighted linear regression may be useful. Overall, the method $\mathrm{lin}_w$ exhibits the greatest robustness across all measurement data resolutions, with the lowest mean estimation error for the entire considered range.

\section{Conclusions}
\label{sec:outlook}
The paper presented methods for estimation of total impedances of branches in systems with sparse measurement setups. The different regression methods discussed allowed the estimation of the total impedances based on measurements of voltage and current magnitudes at both ends of the branch. The methods were compared in simulation studies on LV distribution feeders with varying number of grid segments and unmeasured intermediate loads, showing that the accuracy of the estimates depends on the specific scenario and estimation method. Furthermore, it was shown that the accuracy decreased when the number of unmeasured segments increases or the time resolution $\Delta t$ of measurements increases. The comparison of the proposed methods with respect to their estimation accuracy resulted in the weighted linear regression being the best method. Hence, the regression method can be used in future research to derive surrogate models that only use limited measured grid nodes for state estimation purposes. Additionally, data from typical line types and geographic information systems can be used to improve the estimation accuracy.

\bibliographystyle{IEEEtran}
\bibliography{bibliography}
\end{document}